\documentclass{elsart}  %{article}
\usepackage{epsf,graphics,natbib}
\begin{document}
\def\fdg{\hbox{$.\!\!^\circ$}}
\def\mycaption#1{\begin{quote}\noindent{#1}\end{quote}}
%\citestyle{aa}
\def\dsp{\def\baselinestretch{2.0}\large\normalsize}
%\dsp

\begin{center} 
{\bf \Large
The Cosmic Microwave Background Radiation
}

\vspace{0.1 in}
{\large
Eric Gawiser\footnote{current address:  Center for Astrophysics and Space Sciences, University of California at San Diego, La Jolla, CA  92037}}
\\
{\normalsize
 Department of Physics, University of
California, Berkeley, CA 94720
}\\
\vspace{0.1 in}
\large
and\\
Joseph Silk\\
\normalsize 
Department of Physics, Astrophysics, 1 Keble Road, University of Oxford, OX1
3NP, UK\\
and\\ 
Departments of Physics and Astronomy and Center for Particle 
Astrophysics, University of
California, Berkeley, CA 94720\\

\vspace{0.2in}

\end{center}

\begin{center}  \bf Abstract
\end{center}

We summarize the theoretical and observational status of the study of 
the Cosmic Microwave Background radiation.  Its thermodynamic spectrum is 
a robust prediction of the Hot Big Bang cosmology and has been 
confirmed observationally.  There are now 76 observations of 
Cosmic Microwave Background anisotropy, which we present in a table 
with references.  We discuss the theoretical origins 
of these anisotropies and explain the standard jargon associated with 
their observation.

\section{Origin of the Cosmic Background Radiation}

Our present understanding of the beginning of the universe is based upon the 
remarkably successful theory of the Hot Big Bang.  We believe that our universe began about 15 billion years ago  as a hot, dense, nearly uniform sea of radiation 
a minute fraction of its present size (formally an infinitesimal
singularity).
If inflation occurred in the first fraction of a second, the universe became 
matter dominated while expanding exponentially and then returned to 
radiation domination by the reheating caused by the decay of the inflaton.  
Baryonic
matter formed within the first second, 
and the nucleosynthesis of the lightest elements 
took only a few minutes as the universe expanded and cooled.  
The baryons were in the form of plasma 
until about 300,000 years after the Big Bang, when the universe had 
cooled to a temperature near 3000 K, sufficiently cool for 
protons to capture free electrons and form atomic hydrogen; this process 
is referred to as recombination.  The recombination epoch 
occurred at a redshift of 1100, meaning that 
the universe has grown over a thousand times larger since then.
The ionization energy of a hydrogen atom is 13.6 eV, but 
recombination did not 
occur until the universe had cooled to a characteristic temperature (kT) of
0.3 eV \citep{padmanabhan93}.  
This delay had several causes.  The high entropy of the universe
made the rate of electron capture only marginally faster than the rate of 
photodissociation.  Moreover, each electron captured directly into the ground
state emits a photon capable of ionizing another newly formed atom, so it 
was through 
recombination into excited states and the cooling of the universe to
temperatures below the ionization energy of hydrogen
that neutral matter finally condensed out of the plasma.  
Until recombination,
the universe was opaque to electromagnetic radiation due to scattering 
of the photons by free electrons.   As recombination occurred, the 
density of free electrons diminished greatly, leading to the decoupling of
matter and radiation as the universe became transparent to light.  

The Cosmic Background Radiation (CBR) 
released during this era of decoupling has a mean free path long 
enough to travel almost unperturbed until the present day, where we 
observe it peaked in the microwave region of the spectrum as the 
Cosmic Microwave Background (CMB).  
We see this radiation today coming from the surface of last 
scattering (which is really a spherical shell of finite thickness) 
at a distance of 
 nearly 15 billion light years.  
This Cosmic Background Radiation was predicted by the Hot Big Bang theory
and discovered at an antenna temperature of 3K 
in 1964 by \citet{penziasw65}.  
The number density of photons in the universe at a redshift $z$ is given by \citep{peebles93} 

\begin{equation}
 n_{\gamma} = 420 (1 + z)^{3} cm ^{-3}  
\end{equation}

\noindent 
where $(1 + z)$ is the factor by which the linear scale of the 
universe has expanded since then.  
The radiation temperature of the universe is given by $T = T_{0} (1 + z)$ so it
is easy to see how the conditions in the early universe 
at high redshifts were hot and dense.  

The CBR is our best probe into the conditions of the early universe.  Theories
of the formation of large-scale structure 
predict the existence of slight inhomogeneities in the distribution of 
matter in the early universe which underwent gravitational 
collapse to form galaxies, galaxy clusters, and superclusters.  These density 
inhomogeneities lead to temperature anisotropies in the CBR 
due to a combination of intrinsic temperature fluctuations and 
gravitational blue/redshifting of the photons leaving under/overdense 
regions.  
The DMR (Differential Microwave Radiometer) 
instrument of the Cosmic Background Explorer (COBE) satellite 
discovered primordial temperature 
fluctuations on angular scales larger than $7^\circ$
of order $\Delta T/T = 10^{-5}$ \citep{smootetal92}.  
Subsequent observations of the CMB have 
revealed temperature anisotropies on smaller 
angular scales which correspond to the physical scale of 
observed structures such as galaxies 
and clusters of galaxies.

\subsection{Thermalization}

There were three main processes by which this radiation interacted with matter 
in the first few hundred thousand years:  Compton scattering, double Compton 
scattering, and thermal bremsstrahlung.
The simplest interaction of matter and radiation is Compton
scattering of a single photon off a free electron, 
$ \gamma + e^{-} \rightarrow \gamma + e^{-}$. 
The photon will transfer
momentum and energy to the electron if it has significant energy in the
electron's rest frame.  However, the scattering will be
well approximated by Thomson scattering if the photon's energy in 
the rest frame of the electron is significantly less than the rest mass, 
$h \nu \ll m_{e}c^{2}$.
When the electron is relativistic, the photon is blueshifted by 
roughly a factor 
$\gamma$ in energy when viewed from the 
electron rest frame, is then emitted at almost the same energy in the 
electron rest frame, and is blueshifted by another factor of $\gamma$
when retransformed to the observer's frame.  Thus, energetic 
electrons can efficiently transfer energy 
to the photon background of the universe.  
This process is referred to as Inverse Compton scattering.
The combination of cases where the photon gives energy to the electron 
and vice versa allows Compton scattering to generate thermal equilibrium 
(which is impossible in the Thomson limit of elastic scattering).
Compton scattering conserves the number of photons.
There exists a similar process, double Compton scattering,
 which produces (or absorbs)
photons, $e^- + \gamma \leftrightarrow e^{-} + \gamma + \gamma $.

Another electromagnetic interaction which occurs in the plasma of the early
universe is Coulomb scattering.  Coulomb scattering establishes 
and maintains thermal equilibrium among the baryons of the photon-baryon 
fluid without affecting the
photons.  However, when electrons encounter ions they experience an 
acceleration and therefore emit electromagnetic radiation.  This is called 
thermal bremsstrahlung or free-free emission.  For an ion $X$, 
we have $e^{-} + X \leftrightarrow e^{-} + X + \gamma$.  The interaction
can occur in reverse because of the ability of the charged particles
to absorb incoming photons; this is called free-free absorption.  Each charged
particle emits radiation, but the acceleration is proportional to the mass,
so we can usually view the electron as being accelerated in the fixed Coulomb
field of the much heavier ion.  
Bremsstrahlung is dominated by electric-dipole 
radiation \citep{shu91} and can also 
produce and absorb photons.  

The net effect is that Compton scattering is dominant
for temperatures above 90 eV whereas bremsstrahlung is the primary process
between 90 eV and 1 eV.  At temperatures above 1 keV, double Compton 
is more efficient than bremsstrahlung.  All three processes occur faster than 
the expansion of the universe and therefore have an impact until decoupling.
A static solution for Compton scattering 
is the Bose-Einstein distribution, 

\begin{equation}
 f_{BE} = \frac {1} {e^{x + \mu} - 1} 
\end{equation}

\noindent 
where $\mu$ is a dimensionless chemical potential \citep{hu95}.  
At high optical depths, Compton scattering can exchange enough energy to 
bring the photons to this Bose-Einstein equilibrium distribution.  A Planckian
spectrum corresponds to zero chemical potential, which will occur only when 
the number of photons and total energy are in the same proportion as they 
would be for a blackbody.  Thus, unless the photon number starts out exactly
right in comparison to the total energy in radiation in the universe, Compton
scattering will only produce a Bose-Einstein distribution
and not a blackbody spectrum.  It is important to note, however, that 
Compton scattering will preserve a 
Planck distribution,

\begin{equation}
 f_{P} = \frac {1}{e^{x} - 1 }. 
\end{equation}

All three interactions
will preserve a thermal spectrum if one is achieved at any point.  It has
long been known that the expansion of the universe serves to decrease
the temperature of a blackbody spectrum, 

\begin{equation}
B_{\nu} = \frac{2 h \nu^{3} / c^{2}} {e^{h \nu / k T} - 1}, 
\end{equation}

\noindent
 but keeps it thermal 
\citep{tolman34}.  This
occurs because both the frequency and temperature decrease as $(1 + z)$
leaving $h \nu / k T$ 
unchanged during expansion.  Although
Compton scattering alone cannot produce a Planck distribution, such a 
distribution will remain unaffected by electromagnetic interactions or the 
universal expansion once it is achieved.  
A non-zero
chemical potential will be reduced to zero by double Compton scattering
and, later, bremsstrahlung which will create and absorb photons until the 
number density matches the energy and a thermal distribution of zero
chemical potential is achieved. 
 This results in the thermalization 
of the CBR at redshifts much greater than that of recombination.

Thermalization, of course, should only be able to create an 
equilibrium temperature over regions that are in causal contact.  
The causal horizon at the time of last scattering was relatively small, 
corresponding to a scale today of about 200 Mpc, or a region of 
angular extent of one degree on the sky.  However, observations of the 
CMB show that it has an isotropic temperature on the sky to the 
level of one part in one hundred thousand!  This is the origin of the 
Horizon Problem, which is that there is no physical mechanism expected 
in the early universe which can produce thermodynamic equilibrium on 
superhorizon scales.  The inflationary universe paradigm 
\citep{guth81,linde82,albrechts82}
solves the Horizon 
Problem by postulating that the universe underwent a brief phase of 
exponential expansion during the first second after the Big Bang, during 
which our entire visible Universe expanded out of a region small 
enough to have already achieved thermal equilibrium.

\section{CMB Spectrum}

\begin{figure}
\begin{center}
\rotatebox{0}{\scalebox{0.75}{\includegraphics{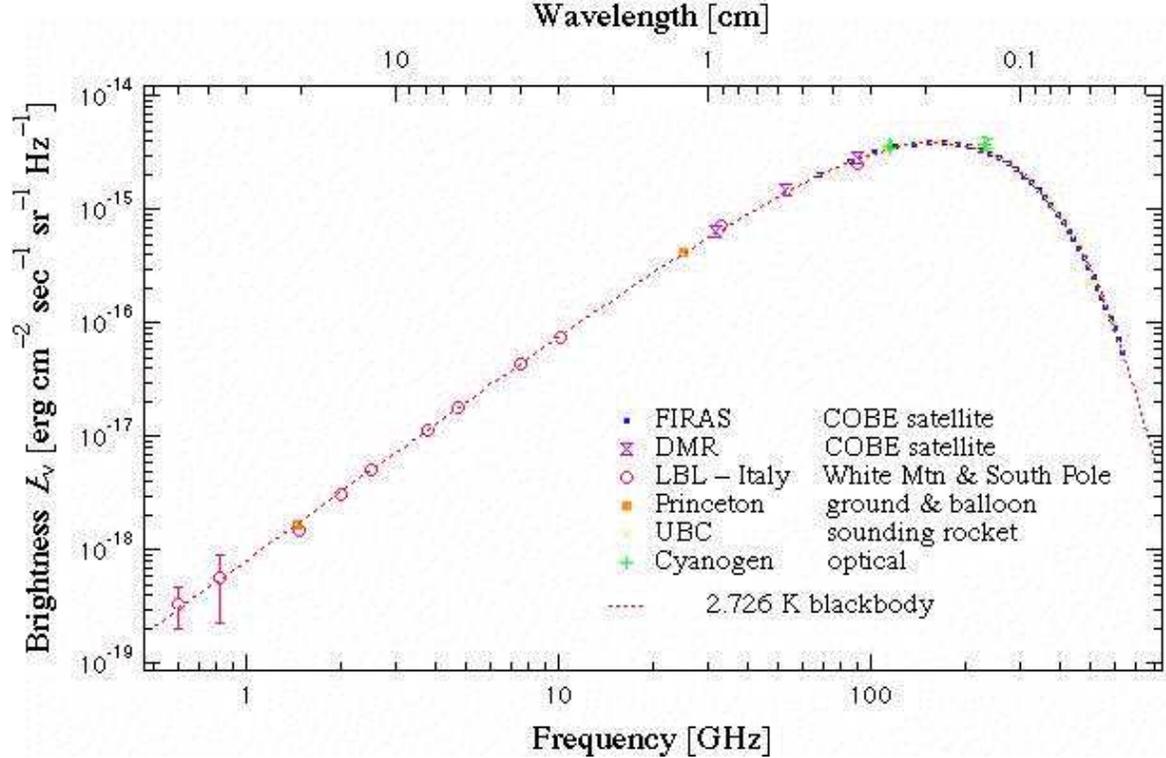}}}
\caption{\small Measurements of the CMB spectrum.}
\label{fig:spectrum}
\end{center}
\end{figure}

The CBR 
is the most perfect blackbody ever
seen, according to the FIRAS (Far InfraRed Absolute 
Spectrometer) instrument of COBE, which measured a temperature 
of $T_0 = 2.726 \pm 0.010$ K \citep{matheretal94}.   
The theoretical prediction that the CBR will have a blackbody spectrum 
appears to be confirmed by the FIRAS observation
 (see Figure \ref{fig:spectrum}).  
But this is not the 
end of the story.  FIRAS only observed the peak of the blackbody.
Other experiments have mapped out the Rayleigh-Jeans part of the
spectrum at low frequency.  Most are consistent with a 2.73 K blackbody, but
some are not.  It is in the low-frequency limit that the greatest spectral
distortions might occur because a Bose-Einstein distribution differs from 
a Planck distribution there.  However, double Compton and 
bremsstrahlung are most effective at low frequencies so 
strong deviations
from a blackbody spectrum are not generally expected.

Spectral distortions in the Wien tail of the spectrum are quite difficult
to detect due to the foreground signal from interstellar dust at those high frequencies.  For example, 
broad emission lines from electron capture at recombination are predicted
in the Wien tail but cannot be distinguished due to foreground 
contamination \citep{whitess94}.  However, because the energy generated 
by star formation and active galactic nuclei is absorbed by interstellar 
dust in all galaxies and then re-radiated in the far-infrared, we expect 
to see an isotropic Far-Infrared Background (FIRB) which dominates the CMB at 
frequencies above a few hundred GHz.  This FIRB has now been detected in 
FIRAS data \citep{pugetetal96,buriganap98,fixsenetal98} and in data 
from the COBE DIRBE instrument \citep{schlegelfd98, dweketal98}.  

%\begin{figure}
%\vbox{%
%\begin{center}
%\leavevmode
%\hbox{%
%angle=-90  %what command does that??
%\epsfxsize=7.5cm
%\epsffile{firas.ps}}  %spectrum2 is portrait, 3 is landscape
%\begin{small}
%% was \figcaption
%\caption{\small Measurements of the CMB spectrum.}
%\end{small}
%\label{fig:spectrum}
%\end{center}}
%\end{figure}

%\begin{figure}
%\centerline{\psfig{file=spectrum.ps,width=4in,angle=-90}}
%\caption{Measurements of the CMB spectrum.}
%\label{fig:spectrum}
%\end{figure}

Although Compton, double Compton, and bremsstrahlung interactions occur 
frequently until decoupling, the complex interplay between them required
to thermalize the CBR spectrum is ineffective at redshifts below 
$10^7$.  
This means that any process after that time
which adds a significant portion of 
energy to the universe  
will lead to a spectral distortion today.  Neutrino decays during this 
epoch should lead to a Bose-Einstein rather than a Planck distribution, and 
this allows the FIRAS observations to set constraints on the decay of 
neutrinos and other particles 
in the early universe \citep{kolbt90}.  The apparent
impossibility of thermalizing 
radiation at low redshift makes the blackbody nature of the CBR strong 
evidence that it did originate in the early universe and as a result serves
to support the Big Bang theory.  

The process of Compton scattering can cause spectral distortions if it
is too late for double Compton and bremsstrahlung to be effective.  In general,
low-frequency photons will be shifted to higher frequencies, thereby 
decreasing the number of photons in the Rayleigh-Jeans region and enhancing
the Wien tail.  This is referred to as a Compton-{\it y} distortion and it
is described by the parameter

\begin{equation}
 y = \int \frac {T_{e}(t)} {m_{e}} \sigma n_{e}(t) dt. 
\end{equation}

\noindent
The apparent temperature drop in the long-wavelength limit is

\begin{equation}
 \frac {\delta T}{T} = - 2 y. 
\end{equation}

\noindent
The most important example of this is Compton scattering of photons off 
hot electrons in galaxy clusters, called the Sunyaev-Zel'dovich (SZ) 
effect.  The electrons transfer energy to the photons, and the spectral
distortion results from the sum of all of the scatterings off 
electrons in 
thermal motion, each of which has a Doppler shift.  The 
SZ effect from clusters can yield a distortion of $y \simeq 10^{-5} - 
10^{-3}$ and these distortions have been observed in several
rich clusters of galaxies.  The FIRAS observations place 
a constraint on any full-sky Comptonization by limiting the average 
$y$-distortion to $y < 2.5 \times 10^{-5}$ \citep{hu95}.  The 
integrated $y$-distortion predicted from the SZ effect of 
galaxy clusters and large-scale structure is over a factor of ten lower 
than this observational constraint \citep{refregiersh98} 
but that from ``cocoons'' of radio galaxies \citep{yamadass99}
 is predicted to be of the same order.  A kinematic SZ effect is caused by 
the bulk velocity of the cluster; this is a small effect which is very 
difficult to detect for individual clusters but will likely 
be measured statistically by the Planck satellite.

\section{CMB Anisotropy}

The temperature anisotropy at a point on the sky $(\theta,\phi)$ can be 
expressed in the basis of spherical harmonics as 
\begin{equation}
 \frac{\Delta T}{T} (\theta, \phi) = 
\sum_{\ell m} a_{\ell m} Y_{\ell m}(\theta, \phi).
\end{equation}

\noindent
A cosmological model predicts the variance 
of the $a_{\ell m}$ coefficients 
over an ensemble of universes (or an ensemble of observational 
points within one universe, if the universe is ergodic).  The
assumptions of rotational symmetry and Gaussianity allow
 us to express this ensemble 
average in terms of the multipoles $C_{\ell}$ as 
\begin{equation}
 \langle a^{*}_{\ell m} a_{\ell' m'} \rangle \equiv 
C_{\ell} \delta_{\ell' \ell} \delta_{m' m}.
\end{equation}

\noindent
The predictions 
of a cosmological model can be expressed in terms of $C_{\ell}$ alone if 
that model predicts a Gaussian distribution of density perturbations, 
in which case the 
$a_{\ell m}$ will have mean zero  
and variance $C_\ell$.

The temperature anisotropies of the CMB detected by COBE are believed 
to result from inhomogeneities in the distribution of 
matter at the epoch of recombination.  Because Compton scattering is an 
isotropic process in the electron rest frame, any primordial anisotropies (as 
opposed to inhomogeneities) should have been smoothed out before
decoupling.  This lends credence
to the interpretation of the observed 
anisotropies as the result of density perturbations
which seeded the formation of galaxies and clusters.  
The discovery of temperature anisotropies by 
COBE provides evidence that 
such density inhomogeneities existed in the early
universe, perhaps caused by 
quantum fluctuations in the scalar field of inflation
or by topological defects resulting from a phase transition 
(see \citealp{kamionkowskik99} for a detailed review of inflationary 
and defect model predictions for CMB anisotropies).    
Gravitational collapse of 
these primordial density inhomogeneities 
appears to have formed
the large-scale structures of galaxies,
clusters, and superclusters that we observe today.  

On large (super-horizon) scales, the anisotropies seen in the CMB are produced 
by the Sachs-Wolfe effect \citep{sachsw67}.  

\begin{equation}
\left( \frac{\Delta T }{T}\right )_{SW} = {\bf v \cdot e}|^e_o -
\Phi|^e_o + \frac{1}{2} \int^e_o h_{\rho \sigma , 0} n^\rho n^\sigma d \xi , 
\end{equation}

where  
the first term is the net 
Doppler shift of the photon 
due to the relative motion of emitter and observer, which 
is referred to as the 
kinematic dipole.  This dipole, first observed by \citet{smootgm77},  
is much larger than other CMB anisotropies and is believed to reflect the 
motion of the Earth relative to the average reference frame of the CMB.  
Most of this motion is due to the peculiar velocity of the 
Local Group of galaxies.  
 The second term represents the gravitational redshift due to a 
difference in 
gravitational potential between the site of photon emission and the 
observer.  The third term is called the Integrated 
Sachs-Wolfe (ISW) effect and is caused by a non-zero time derivative of 
the metric along the photon's path of travel due to potential decay, 
gravitational waves, or non-linear structure evolution (the Rees-Sciama 
effect).  In a matter-dominated universe with scalar density 
perturbations the integral vanishes on linear scales.  
This equation gives the redshift 
from emission to observation, but there is also an intrinsic $\Delta T/T$ 
on the last-scattering surface due to the local density of photons.  
For adiabatic perturbations, we have \citep{whiteh97} an intrinsic 

\begin{equation}
\frac{\Delta T }{T} = \frac{1}{3} \frac{\delta \rho}{\rho} = 
 \frac{2}{3} \Phi .  
\end{equation} 

Putting the observer at $\Phi=0$ (the observer's gravitational 
potential merely adds a constant energy to all CMB photons) 
this leads to a net Sachs-Wolfe effect 
of $\Delta T /T = - \Phi/3$ which means that overdensities lead 
to cold spots in the CMB.

\subsection{Small-angle anisotropy}

Anisotropy measurements on 
small angular scales ($0\fdg1$ to $1^{\circ}$) 
are expected to reveal the so-called
first acoustic
 peak of the CMB power spectrum.  This peak in the anisotropy power spectrum
corresponds 
to the scale where acoustic oscillations of the photon-baryon fluid caused
by primordial density inhomogeneities are just reaching their maximum 
amplitude at the surface of last scattering i.e. the sound horizon 
at recombination.  Further acoustic  
peaks occur at scales that are reaching their second, third, fourth, etc.
antinodes of oscillation.

\begin{figure}
\begin{center}
\scalebox{0.75}{\includegraphics{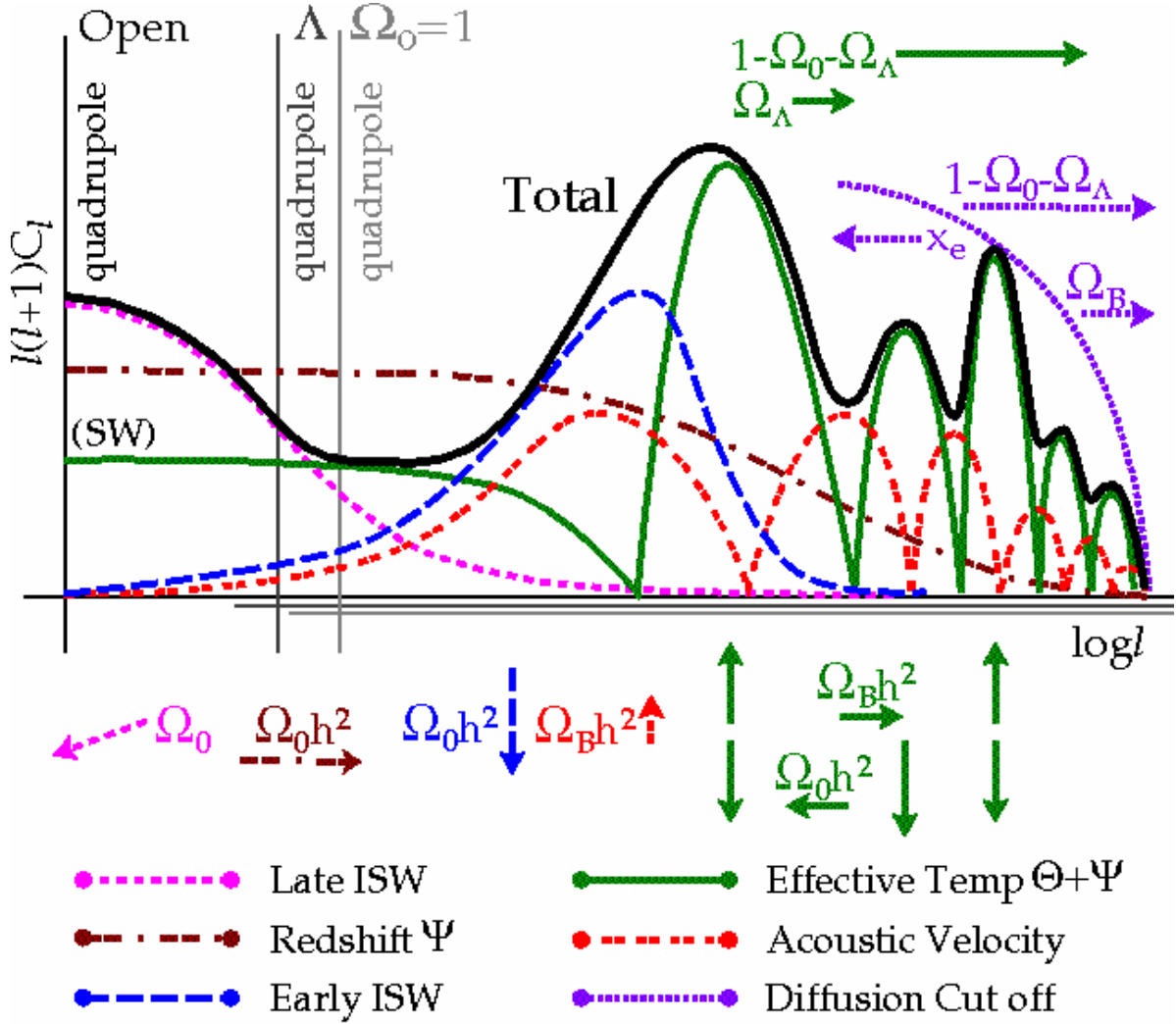}}
\caption{\small
Dependence of CMB anisotropy power spectrum on 
cosmological parameters.}
\label{fig:hu}
\end{center}
\end{figure}

%\begin{figure}
%\centerline{\psfig{file=hu_fig5.ps,width=4in}}
%\caption{Dependence of CMB anisotropy power spectrum on 
%cosmological parameters.}
%\label{fig:hu}
%\end{figure}

Figure \ref{fig:hu} (from \citealp{huss97})
%Hu, Sugiyama, \& Silk 1995) 
shows the dependence of the CMB anisotropy power spectrum on a number of cosmological parameters.  The acoustic oscillations in 
density (light solid line) are sharp 
here because they are really being plotted against spatial scales, which 
are then smoothed as they are projected through the last-scattering surface 
onto angular scales.  The troughs in the density oscillations are filled in 
by the 90-degree-out-of-phase velocity oscillations (this is a Doppler 
effect but does not correspond to the net peaks, which
are best referred to as acoustic peaks rather than Doppler peaks).  
The origin of this plot
 is at a different place for different values of the matter density and the cosmological constant; the negative 
spatial curvature of an open universe makes a given spatial scale 
correspond to a smaller angular scale.
The Integrated Sachs-Wolfe (ISW) effect occurs whenever 
gravitational potentials decay due to a lack of matter dominance.  Hence 
the early ISW effect occurs just after recombination when the density 
of radiation is still considerable and serves to broaden the first 
acoustic peak at scales just larger than the horizon size at 
recombination.  And for a present-day matter density less than critical, 
there is a late ISW effect that matters on very large angular scales - it 
is greater in amplitude for open universes than for lambda-dominated 
because matter domination ends earlier in an open universe for the same 
value of the matter density today.  The late ISW effect should correlate with 
large-scale structures that are otherwise detectable at $z \sim1$, and this 
allows the CMB to be cross-correlated with observations of the X-ray background to determine $\Omega$ 
\citep{crittendent96, kamionkowski96, boughnct98, kamionkowskikink99}
 or with observations 
of large-scale structure to determine the bias of galaxies 
\citep{suginoharass98}.

For a given model,  
the location of the 
first acoustic peak can yield information about $\Omega$, the ratio of 
the density of the universe to the critical density needed to stop
its expansion.  For adiabatic density perturbations, the first 
acoustic peak will occur at $\ell = 220 \Omega^{-1/2}$
\citep{kamionkowskiss94}.  The ratio of $\ell$ values of the peaks 
is a robust test of the nature of the density perturbations; for 
adiabatic perturbations these will have ratio 1:2:3:4 whereas for 
isocurvature perturbations the ratio should be 1:3:5:7 \citep{huw96}. 
A mixture of adiabatic and isocurvature 
perturbations is possible, and this test 
should reveal it.

As illustrated in Figure \ref{fig:hu}, the amplitude of the 
acoustic peaks depends on the baryon fraction $\Omega_b$, 
the matter density $\Omega_0$, and Hubble's constant 
$H_0 = 100 h$ km/s/Mpc.   
A precise measurement of all three acoustic peaks can reveal 
the fraction of hot dark matter and even potentially the number 
of neutrino species \citep{dodelsongs96}.  
Figure \ref{fig:hu} shows the envelope of the 
CMB anisotropy damping tail on arcminute scales, 
where the fluctuations are decreased due to photon diffusion 
\citep{silk67} as well as the finite thickness of the last-scattering 
surface.  This damping tail
 is a sensitive probe 
of cosmological parameters and has the potential to 
break degeneracies between models which explain the larger-scale anisotropies
\citep{huw97a,metcalfs98}.  The characteristic angular 
scale for this damping is given by
 $1.8' \Omega_B^{-1/2} \Omega_0^{3/4} h^{-1/2}$ \citep{whitess94}.  

	There is now a plethora of theoretical 
models which predict the development of primordial 
density perturbations into microwave 
background anisotropies.  These models differ in their explanation 
of the origin of density inhomogeneities 
(inflation or topological defects), the nature of the dark matter (hot,
cold, baryonic, or a mixture of the three), 
the curvature of the universe ($\Omega$), 
the value of the cosmological constant ($\Lambda$), 
the value of Hubble's constant, and 
the possibility of reionization which 
wholly or partially erased temperature anisotropies in the CMB on 
scales smaller than the horizon size.  Available data does not allow 
us to constrain all (or even most) of these parameters, so analyzing 
current CMB anisotropy data requires
a model-independent approach.  It seems
reasonable to view the mapping of the acoustic peaks as
a means of determining the nature of parameter space
before going on to fitting cosmological parameters directly. 

\subsection{Reionization}

The possibility 
that post-decoupling interactions between ionized matter and the CBR
have affected the anisotropies on scales 
smaller than those measured by COBE is of great significance
for current experiments.  
Reionization is inevitable but its effect on anisotropies 
depends significantly on when it occurs
(see \citealp{haimank99} for a review).  
  Early reionization leads to 
a larger optical 
depth and therefore a greater damping of the anisotropy 
power spectrum due to the secondary scattering of CMB photons 
off of the newly free electrons.    
For a universe with critical matter density and constant ionization fraction 
$x_e$, the optical depth as a function of redshift is given by 
\citep{whitess94}

\begin{equation}
\tau \simeq 0.035 \Omega_B h x_e z^{3/2},
\end{equation}

which allows us to determine the redshift of reionization $z_\ast$ at 
which $\tau = 1$, 

\begin{equation}
z_\ast \simeq 69 \left(\frac{h}{0.5}\right)^{-\frac{2}{3}} 
\left(\frac{\Omega_B}{0.1}\right)^{-\frac{2}{3}}x_e^{-\frac{2}{3}}
\Omega^{\frac{1}{3}}, 
\end{equation}

where the scaling with $\Omega$ applies to an open universe only.  
At scales smaller than the horizon size at reionization, 
$\Delta T/T$ is reduced by the factor $e^{-\tau}$.  

Attempts to measure
the temperature anisotropy on angular scales of less than a degree which 
correspond to the size of galaxies could have led to a surprise; 
if the universe was reionized after recombination to the extent
that the CBR was significantly scattered 
at redshifts less than 1100, the small-scale 
primordial anisotropies would have been washed out.
To have an appreciable optical depth 
for photon-matter interaction, reionization cannot have occurred 
much later than a redshift of 20 \citep{padmanabhan93}.  
Large-scale anisotropies such as those
seen by COBE are not expected to be affected by reionization because they 
encompass regions of the universe which were not yet in causal contact
even at the proposed time of reionization.  However, the apparently high 
amplitiude of degree-scale anisotropies is a strong argument against the 
possibility of early ($z\geq50$) reionization.  
On arc-minute scales, the 
interaction of photons with reionized matter is expected to have eliminated
the primordial anisotropies and replaced them with smaller secondary 
anisotropies from this new surface of last scattering (the 
Ostriker-Vishniac effect and patchy reionization, see next section). 

\subsection{Secondary Anisotropies}

Secondary CMB anisotropies 
occur when the photons of the Cosmic Microwave 
Background radiation are scattered after the original last-scattering 
surface (see \citealp{refregier99} for a review).  
  The shape of the blackbody 
spectrum can be altered through 
inverse Compton scattering by the thermal Sunyaev-Zel'dovich (SZ) effect
\citep{sunyaevz72}.
  The effective temperature of 
the blackbody can be shifted locally by 
a doppler shift from the peculiar velocity of the scattering medium (the 
kinetic SZ and Ostriker-Vishniac effects, \citealp{ostrikerv86}) 
as well as by passage through
the changing gravitational potential caused by the 
collapse of nonlinear structure (the Rees-Sciama effect, 
\citealp{reess68}) or 
the onset of curvature or cosmological constant domination (the Integrated 
Sachs-Wolfe effect).  
Several simulations of the impact of 
patchy reionization have been performed 
\citep{aghanimetal96, knoxsd98, gruzinovh98, peeblesj98}.
%(Aghanim et al. 1996, 
%Knox, Scoccimarro, \& Dodelson 1998, Gruzinov \& Hu 1998, Peebles \& 
%Juskiewicz 1998).  
The SZ effect itself is independent of redshift, so it can yield 
information on clusters at much higher redshift than does X-ray 
emission.  However, nearly all clusters are unresolved for $10'$ resolution 
so higher-redshift clusters occupy less of the beam and therefore their SZ
effect is in fact dimmer.  In the 4.5$'$ channels of Planck this will 
no longer be true, and the SZ effect can 
probe cluster abundance at high redshift.  An additional 
secondary anisotropy is that caused by gravitational lensing (see e.g. 
\citealp{cayonms93, cayonms94, metcalfs97, mgsc97}).  
Gravitational lensing imprints 
slight non-Gaussianity in the CMB from which it might be possible  
to determine the matter power spectrum 
\citep{seljakz98, zaldarriagas98b}.

\subsection{Polarization Anisotropies}	

Polarization of the Cosmic Microwave Background radiation 
\citep{kosowsky94, kamionkowskiks97, zaldarriagas97}
arises 
due to local quadrupole anisotropies at each point on the surface 
of last scattering (see \citealp{huw97b} for a review).
  Scalar (density) perturbations generate curl-free 
(electric mode) polarization only, but tensor (gravitational wave) 
perturbations can generate divergence-free (magnetic mode) polarization.  
Hence the polarization of the CMB is a potentially useful probe of 
the level of gravitational waves in the early universe
\citep{seljakz97, kamionkowskik98}, especially 
since current indications are that the large-scale primary 
anisotropies seen by COBE do not contain a measurable fraction 
of tensor contributions \citep{gawisers98}.  A thorough review 
of gravity waves and CMB polarization is given by \citet{kamionkowskik99}.

\subsection{Gaussianity of the CMB anisotropies}

The processes turning density inhomogeneities into CMB anisotropies 
are linear, so cosmological models that predict gaussian primordial 
density inhomogeneities also predict a gaussian distribution of 
CMB temperature fluctuations.  Several techniques have been developed 
to test COBE and future datasets for deviations from gaussianity
\citep[e.g.][]{kogutetal96b, ferreiram97, ferreirams97}.  Most 
tests have proven negative, but a few claims of non-gaussianity have 
been made.  \citet{gaztanagafe98} found a very marginal indication 
of non-gaussianity in the spread of results for degree-scale 
CMB anisotropy observations being greater than the expected sample 
variances.  \citet{ferreiramg98} have claimed a detection of non-gaussianity 
at multipole $\ell=16$ using a bispectrum statistic,  
and \citet{pandovf98} find a non-gaussian wavelet coefficient correlation 
on roughly $15^\circ$ scales in the North Galactic hemisphere.  Both 
of these methods produce results consistent with gaussianity, however, if 
a particular area of several pixels is eliminated from the dataset 
\citep{bromleyt99}.   A true sky signal should be larger than several 
pixels so instrument noise is the most likely source of the non-gaussianity. 
  A different area appears to cause each detection, giving 
evidence that the COBE dataset had non-gaussian instrument noise in at 
least two areas of the sky.

\subsection{Foreground contamination}

Of particular concern in measuring CMB anisotropies is the issue of foreground
contamination.
Foregrounds which can affect CMB observations include
galactic radio emission (synchrotron and free-free), galactic infrared
emission (dust), extragalactic radio sources (primarily elliptical galaxies, 
active galactic nuclei, and quasars), extragalactic infrared sources (mostly
dusty spirals and high-redshift 
starburst galaxies), and the Sunyaev-Zel'dovich effect from 
hot gas in 
galaxy clusters.  The COBE team has gone to great lengths to analyze their
data for possible foreground contamination and routinely eliminates everything
within about $30^{\circ}$ of the galactic plane.  

\begin{figure}
\begin{center}
\scalebox{0.75}{\includegraphics{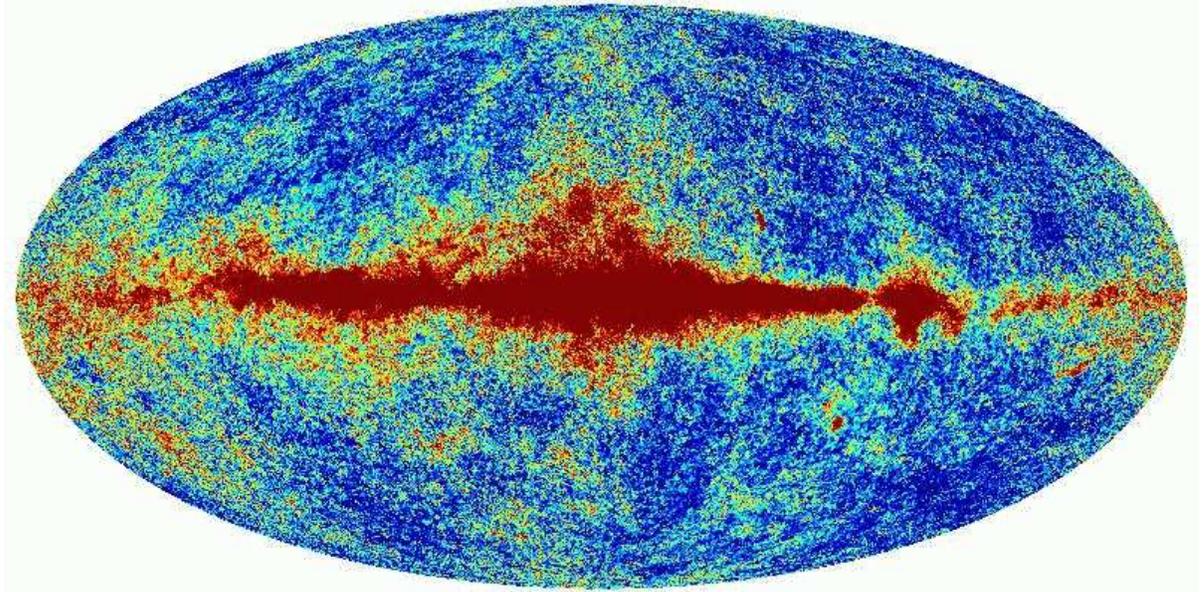}}
\caption{\small 
WOMBAT Challenge simulation of CMB anisotropy map 
that might be observed by the MAP satellite at 90 GHz, 13' resolution, 
containing CMB, instrument noise, and foreground contamination.  The 
resolution is degraded by the pixelization of your monitor or printer.}
\label{fig:total}
\end{center}
\end{figure}

An instrument with large 
resolution such as COBE is most sensitive to the diffuse foreground emission
of our Galaxy, but small-scale anisotropy experiments need to worry  
about extragalactic sources as well.  
Because foreground 
and CMB anisotropies are assumed to be uncorrelated, they should add in 
quadrature, leading to an increase in the measurement of CMB anisotropy 
power.  
Most CMB instruments, however,
 can identify foregrounds by their spectral signature
across multiple 
frequencies or their display of the beam response characteristic
of a point source.  This leads to an attempt at foreground subtraction, 
which can cause an underestimate of CMB anisotropy if some true
signal is subtracted along with the foreground.
Because 
they are now becoming critical, extragalactic foregrounds 
have been studied in detail 
\citep{toffolattietal98, refregiersh98, 
gawisers97, sokasiangs98, gawiserjs98}.  
The Wavelength-Oriented Microwave Background Analysis Team 
(WOMBAT, see \citealp{gawiseretal98, jaffeetal99}) has made Galactic 
and extragalactic foreground predictions and full-sky simulations of 
realistic CMB skymaps containing foreground contamination 
available to the public (see http://astro.berkeley.edu/wombat).  
One of these CMB simulations is shown in Figure \ref{fig:total}.  
\citet{tegmarketal99} used a Fisher matrix analysis to show that 
simultaneously estimating foreground model parameters and cosmological 
parameters can lead to a factor of a few degradation in the precision 
with which the cosmological parameters can be determined by CMB anisotropy 
observations, so foreground prediction and subtraction is likely to be 
an important aspect of future CMB data analysis.  

Foreground contamination may turn out to be a serious problem for 
measurements of CMB polarization anisotropy.  While free-free emission 
is unpolarized, synchrotron radiation displays a linear polarization 
determined by the coherence of the magnetic field along the 
line of sight; this is typically on the order 
of 10\% for Galactic synchrotron and between 5 and 10\% for flat-spectrum 
radio sources.
The CMB is  
expected to show a large-angular scale linear polarization of about 10\%, 
so the prospects for detecting polarization anisotropy are no worse than 
for temperature anisotropy although 
higher sensitivity is required.  
However, the small-angular scale electric mode of linear 
polarization which is a probe of several cosmological parameters 
and the magnetic mode that serves as a probe 
of tensor perturbations are expected to have much lower amplitude and 
may be swamped by foreground polarization.  
Thermal and spinning dust grain emission can also be polarized.  
It may turn out that dust emission is the only significant source 
of circularly polarized microwave photons since the CMB cannot have 
circular polarization.

%\begin{figure}
%\centerline{\psfig{file=total.ps,width=5in}}
%\caption{Compilation of CMB Anisotropy observations.}   
%\mycaption{WOMBAT Challenge simulation of CMB anisotropy map 
%that may be observed by the MAP satellite at 90 GHz, 13' resolution, 
%containing CMB, instrument noise, and foreground contamination.
%}
%\label{fig:total}
%\end{figure}

\section{Cosmic Microwave Background Anisotropy Observations}
\label{sect:obs_cmb}

Since the COBE DMR detection of CMB anisotropy \citep{smootetal92}, there have
been over thirty additional measurements of anisotropy on angular scales
ranging from $7^{\circ}$ to $0\fdg3$, and upper limits have been set
on smaller scales.  
%Shown in Figure \ref{fig:obs_cmb} are COBE \citep{tegmarkh97},
%FIRS \citep{gangaetal94},
%Tenerife \citep{gutierrezetal97},
%South Pole \citep{gundersenetal94},
%BAM \citep{tuckeretal97},
%ARGO \citep{masietal96},
%Python \citep{cobleetal99, plattetal97}, 
%MAX \citep{limetal96,tanakaetal96},
%MSAM \citep{wilsonetal99},
%SK \citep{netterfieldetal97}, 
%CAT \citep{scottetal96,bakeretal99},
%OVRO/RING \citep{leitchetal98},
%WD \citep{tuckeretal93},
%OVRO \citep{readheadetal89},
%SUZIE \citep{churchetal97},
%ATCA \citep{subrahmanyanetal93},
%and VLA \citep{partridgeetal97}.  

The COBE DMR observations 
were pixelized into a skymap, from which it is possible to analyze any 
particular multipole within the resolution of the DMR.  
Current small angular scale 
CMB anisotropy observations are insensitive to both high $\ell$ and 
low $\ell$
multipoles because they cannot measure features smaller than their
resolution and are insensitive to features larger than the 
size of the patch of sky observed.
The next satellite mission, NASA's 
Microwave Anisotropy Probe 
(MAP), is scheduled for launch in Fall 2000
and will map angular scales down to $0\fdg2$ with high precision over most of 
the sky.  An even more precise satellite, ESA's Planck, is scheduled 
for launch in 2007.  
  Because COBE observed such large angles, the DMR data can only  
constrain the amplitude $A$ and index $n$ of 
the primordial power spectrum in wave number $k$, $P_p(k) = A k^{n}$, and  
these constraints are not tight enough 
to rule out very many classes of cosmological models.

\begin{figure}
\begin{center}
\scalebox{0.75}{\includegraphics{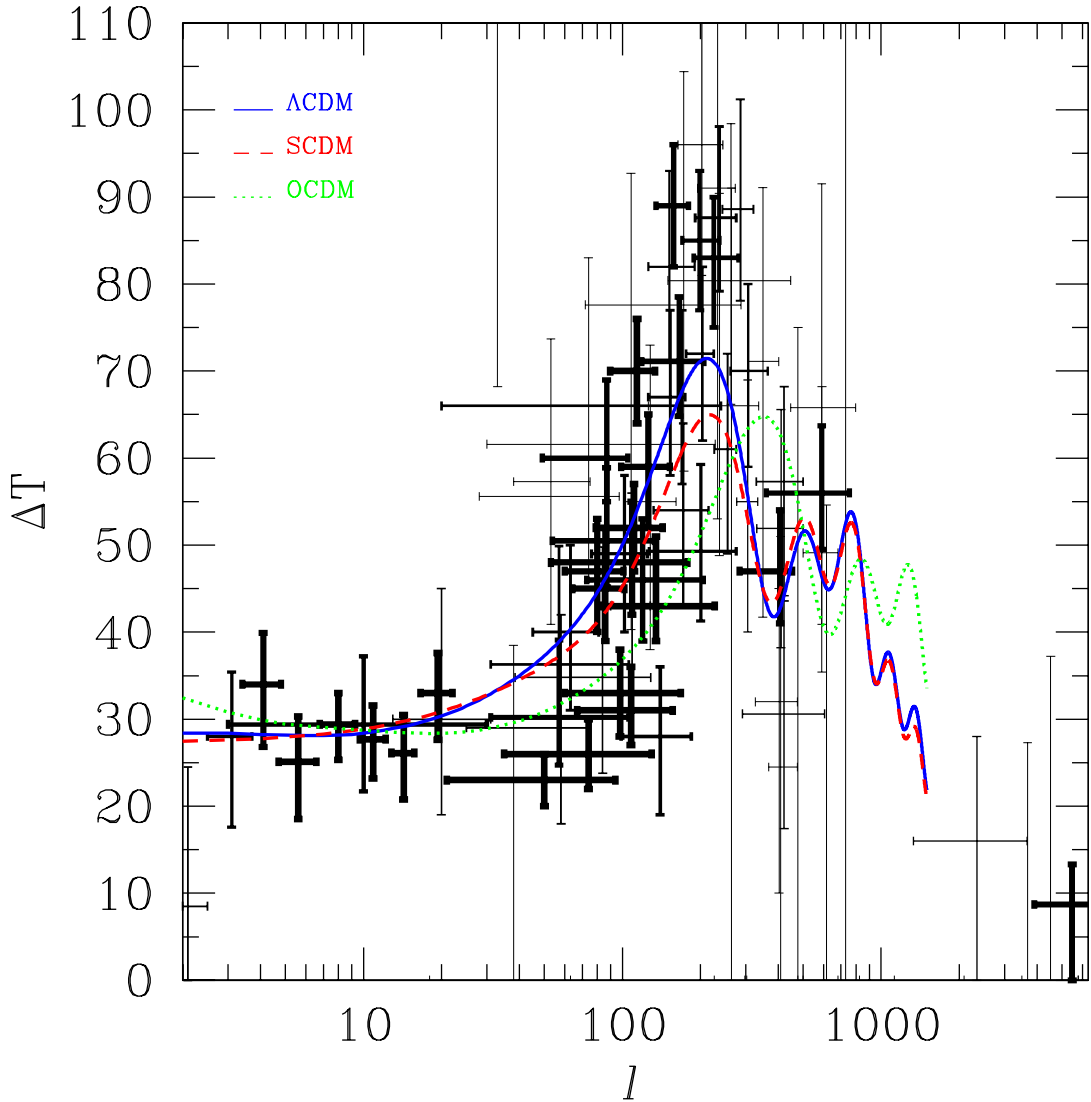}}
\caption{\small Compilation of CMB Anisotropy observations. 
Vertical error bars represent
$1\sigma$ uncertainties  
and horizontal error bars show the range from $\ell_{min}$ to $\ell_{max}$ 
of Table 1.  The line thickness is inversely proportional to the variance 
of each measurement, emphasizing the tighter constraints.  All three models 
are consistent with the upper limits at the far right, but the Open CDM 
model (dotted) is a poor fit to the data, which prefer models 
with an acoustic peak near $\ell=200$ with an amplitude close to that of 
$\Lambda$CDM (solid).}
\label{fig:obs_cmb}
\end{center}
\end{figure}

Until the next satellite is flown, the promise
of microwave background anisotropy measurements to measure 
cosmological parameters rests with a series of ground-based and 
balloon-borne
anisotropy instruments which have already published results (shown 
in Figure \ref{fig:obs_cmb})  
or will report results in 
the next few years (MAXIMA, BOOMERANG, TOPHAT, ACE, MAT, VSA, CBI, DASI, see 
\citealp{leeetal99} and \citealp{halperns99}).
Because they are not satellites, these instruments face the problems of 
shorter observing times and less sky coverage, although significant 
progress has been made in those areas.  They fall into 
three categories:  high-altitude balloons, interferometers, and 
other ground-based instruments.
Past, present, and future balloon-borne instruments are 
FIRS, MAX, MSAM, ARGO, BAM, MAXIMA, QMAP, HACME, 
BOOMERANG, TOPHAT, and ACE.  Ground-based 
interferometers include CAT, JBIAC, SUZIE, BIMA, 
ATCA, VLA, VSA, CBI, and DASI, and other ground-based 
instruments are TENERIFE, SP, PYTHON, SK, OVRO/RING, VIPER, MAT/TOCO,
IACB, and WD.    
Taken as a whole, they have the potential to yield very useful 
measurements of the radiation power spectrum of the CMB on degree and 
subdegree scales.  Ground-based non-interferometers have to discard a large
fraction of data and undergo careful further data reduction to eliminate 
atmospheric contamination.  Balloon-based instruments need to keep a careful 
record of their pointing to reconstruct it during data analysis.  
Interferometers may be the most promising technique at present but they 
are the least developed, and most instruments are at radio frequencies 
and have very narrow frequency 
coverage, making foreground contamination a major concern.  
In order to use small-scale CMB anisotropy measurements to constrain 
cosmological models we need to be confident of their 
validity and to trust the error bars.  This will allow us to discard badly
contaminated data and to give greater weight to the more precise measurements 
in fitting models.  Correlated noise is a great concern for instruments 
which lack a rapid chopping because the $1/f$ noise causes correlations 
on scales larger than the beam 
in a way that can easily mimic CMB anisotropies.
Additional issues are sample variance caused by the combination of 
 cosmic variance and limited sky coverage and foreground contamination.

%[TABLE OF CMB OBSERVATIONS WITH REFERENCES]

{\small
\begin{table}[h] 
%\baselinestretch{1.0}
%\small
\caption{Complete compilation of CMB anisotropy observations 1992-1999, 
with maximum likelihood $\Delta T$, upper and lower 1$\sigma$ uncertainties (not 
including calibration uncertainty),  
the weighted center of the window function, the $\ell$ values where the 
window function falls to $e^{-1/2}$ of its maximum value, the 
1 $\sigma$ calibration uncertainty, and references given below.}
\label{tab:obs}
\begin{center}
\begin{tabular}{|l|l|l|l|l|l|l|l|l|}
\hline
Instrument &  $\Delta T$ ($\mu$K) & +$1\sigma(\mu$K) & -$1\sigma(\mu$K) & $\ell_{eff}$ & 
   $\ell_{min}$ & $\ell_{max}$ & 1$\sigma$ cal. & ref. \\
\hline
COBE1&	8.5&16.0&8.5&2.1&2&2.5&0.7&1\\
COBE2&	28.0&7.4&10.4&3.1&2.5&3.7&0.7&1\\
COBE3&	34.0&5.9&7.2&4.1&3.4&4.8&0.7&1\\
COBE4&	25.1&5.2&6.6&5.6&4.7&6.6&0.7&1\\
COBE5&	29.4&3.6&4.1&8.0&6.8&9.3&0.7&1\\
COBE6&	27.7&3.9&4.5&10.9&9.7&12.2&0.7&1\\
COBE7&	26.1&4.4&5.3&14.3&12.8&15.7&0.7&1\\
COBE8&	33.0&4.6&5.4&19.4&16.6&22.1&0.7&1\\
FIRS&	29.4&7.8&7.7&10&3&30&--$^a$&2\\
TENERIFE&30&15&11&20&13&31&--$^a$&3\\
IACB1&	111.9&49.1&43.7&33&20&57&~20&4\\
IACB2&	57.3&16.4&16.4&53&38&75&~20&4\\
SP91&	30.2&8.9&5.5&57&31&106&~15&5\\
SP94&	36.3&13.6&6.1&57&31&106&15&5\\
BAM& 	55.6&27.4&9.8&74&28&97&20&6\\
ARGO94&	33&5&5&98&60&168&5 &7\\
ARGO96&	48&7&6&109&53&179&10&8\\
JBIAC&	43&13&12&109&90&128&6.6&9\\
QMAP(Ka1)&47.0&6&7&80&60&101&12&10\\
QMAP(Ka2)&59.0&6&7&126&99&153&12&10\\
QMAP(Q)&52.0&5&5&111&79&143&12&10\\
MAX234&	46&7&7&120&73&205&10&11\\
MAX5&	43&8&4&135&81&227&10&12\\
MSAMI&	34.8&15&11&84&39&130&5&13\\
MSAMII&	49.3&10&8&201&131&283&5&13\\
MSAMIII&47.0&7&6&407&284&453&5&13\\
\hline
\end{tabular}
\end{center}
\end{table}
}

{\small
\begin{table}[h] 
%\baselinestretch{1.0}
%\small
\setcounter{table}{0}
%\caption{Table 1 (cont.)}
\label{tab:obs2}
\begin{center}
\begin{tabular}{|l|l|l|l|l|l|l|l|l|}
\hline
\footnotesize
Instrument &  $\Delta T$ ($\mu$K) & +$1\sigma(\mu$K) & -$1\sigma(\mu$K) & $\ell_{eff}$ & 
   $\ell_{min}$ & $\ell_{max}$ & 1$\sigma$ cal. & ref. \\
\hline
PYTHON123&60&9&5&87&49&105&20&14\\
PYTHON3S&66&11&9&170&120&239&20&14\\
PYTHONV1&23&3&3&50&21&94&17$^b$ &15\\
PYTHONV2&26&4&4&74&35&130&17&15\\
PYTHONV3&31&5&4&108&67&157&17&15\\
PYTHONV4&28&8&9&140&99&185&17&15\\
PYTHONV5&54&10&11&172&132&215&17&15\\
PYTHONV6&96&15&15&203&164&244&17&15\\
PYTHONV7&91&32&38&233&195&273&17&15\\
PYTHONV8&0&91&0&264&227&303&17&15\\
SK1$^c$&	50.5&8.4&5.3&87&58&126&11&16\\
SK2&	71.1&7.4&6.3&166&123&196 &11&16\\
SK3&	87.6&10.5&8.4&237&196&266&11&16\\
SK4&	88.6&12.6&10.5&286&248&310&11&16\\
SK5&	71.1&20.0&29.4&349&308&393&11&16\\
TOCO971&40&10&9&63&45&81&10&17\\
TOCO972&45&7&6&86&64&102&10&17\\
TOCO973&70&6&6&114&90&134&10&17\\
TOCO974&89&7&7&158&135&180&10&17\\
TOCO975&85&8&8&199&170&237&10&17\\
TOCO981&55&18&17&128&102&161&8&18\\
TOCO982&82&11&11&152&126&190&8&18\\
TOCO983&83&7&8&226&189&282&8&18\\
TOCO984&70&10&11&306&262&365&8&18\\
TOCO985&24.5&26.5&24.5&409&367&474&8&18\\
VIPER1&	61.6&31.1&21.3&108&30&229&8&19\\
VIPER2&	77.6&26.8&19.1&173&72&287&8&19\\
VIPER3&	66.0&24.4&17.2&237&126&336&8&19\\
VIPER4&	80.4&18.0&14.2&263&150&448&8&19\\
VIPER5&	30.6&13.6&13.2&422&291&604&8&19\\
VIPER6&	65.8&25.7&24.9&589&448&796&8&19\\
\hline
\end{tabular}
\end{center}
\end{table}
}

{\small
\begin{table}[h] 
%\baselinestretch{1.0}
%\small
%\setcounter{table}{0}
%\caption{Table 1 (cont.)}
\label{tab:obs3}
\begin{center}
\begin{tabular}{|l|l|l|l|l|l|l|l|l|}
\hline
\footnotesize
Instrument &  $\Delta T$ ($\mu$K) & +$1\sigma(\mu$K) & -$1\sigma(\mu$K) & $\ell_{eff}$ & 
   $\ell_{min}$ & $\ell_{max}$ & 1$\sigma$ cal. & ref. \\
\hline
BOOM971&29&13&11&58&25&75&8.1&20\\
BOOM972&49&9&9&102&76&125&8.1&20\\
BOOM973&67&10&9&153&126&175&8.1&20\\
BOOM974&72&10&10&204&176&225&8.1&20\\
BOOM975&61&11&12&255&226&275&8.1&20\\
BOOM976&55&14&15&305&276&325&8.1&20\\
BOOM977&32&13&22&403&326&475&8.1&20\\
BOOM978&0&130&0&729&476&1125&8.1&20\\
CAT96I&	51.9&13.7&13.7&410&330&500&10&21\\
CAT96II&49.1&19.1&13.7&590&500&680&10&21\\
CAT99I&	57.3&10.9&13.7&422&330&500&10&22\\
CAT99II&0.&54.6&0.&615&500&680&10&22\\
OVRO/RING&56.0&7.7&6.5&589&361&756&4.3&23\\
HACME&0.&38.5&0.&38&18&63&--$^a$&29\\
WD& 	0.&75.0&0.&477&297&825&30&24\\
SuZIE&	16&12&16&2340&1330&3070&8&25\\
VLA& 	0.&27.3&0.&3677&2090&5761&--$^a$&26\\
ATCA& 	0.&37.2&0.&4520&3500&5780&--$^a$&27\\
BIMA&	8.7&4.6&8.7&5470&3900&7900&--$^a$&28\\
\hline
\end{tabular}
\end{center}
\footnotesize
REFERENCES:  1--\citet{tegmarkh97,kogutetal96c}
2--\citet{gangaetal94}
3--\citet{gutierrezetal99}
4--\citet{femeniaetal98}
5--\citet{gangaetal97b,gundersenetal95}
6--\citet{tuckeretal97}
7--\citet{ratraetal99}
8--\citet{masietal96}
9--\citet{dickeretal99}
10--\citet{docetal98b}
11--\citet{clappetal94, tanakaetal96}
12--\citet{gangaetal98}
13--\citet{wilsonetal99}
14--\citet{plattetal97}
15--\citet{cobleetal99}
16--\citet{netterfieldetal97}
17--\citet{torbetetal99}
18--\citet{milleretal99}
19--\citet{petersonetal99}
20--\citet{mauskopfetal99}
21--\citet{scottetal96}
22--\citet{bakeretal99}
23--\citet{leitchetal98}
24--\citet{ratraetal98}
25--\citet{gangaetal97a,churchetal97}
26--\citet{partridgeetal97}
27--\citet{subrahmanyanetal93}
28--\citet{holzapfeletal99}
29--\citet{starenetal99}

$^a$Could not be determined from the literature.

$^b$Results from combining the +15\% and -12\% 
calibration uncertainty with the 3$\mu$K beamwidth uncertainty.  The 
non-calibration errors on the PYTHONV
datapoints are highly correlated.

$^c$The SK $\Delta T$ and error bars 
have been re-calibrated according to the 
5\% increase recommended by \citet{masonetal99} 
and the 2\% decrease in $\Delta T$ due to foreground contamination 
found by \citet{docetal97}.
\end{table}
}

Figure \ref{fig:obs_cmb}
shows our compilation of CMB anisotropy observations without 
adding any theoretical curves to bias the eye\footnote{This figure and 
our compilation of CMB anisotropy observations are  
available at http://mamacass.ucsd.edu/people/gawiser/cmb.html; CMB 
observations have also been compiled by \citet{smoots98} and
at http://www.hep.upenn.edu/\~{}max/cmb/experiments.html and\\
http://www.cita.utoronto.ca/\~{}knox/radical.html}.  
It is clear that a straight 
line is a poor but not implausible fit to the data. 
There is a clear rise around $\ell=100$ and then a 
drop by $\ell=1000$.  This is not yet good enough to give a clear 
determination of 
 the 
curvature of the universe, 
let alone fit several cosmological parameters.  
However, the current data  prefer 
adiabatic structure formation models over isocurvature 
models \citep{gawisers98}.  
If analysis is restricted to 
adiabatic CDM models, a value of the total density near critical is  
preferred \citep{dodelsonk99}.

\subsection{Window Functions}

The sensitivity of these instruments to various multipoles is called
their window function.  
%Figure 3 shows window functions for the 
%relevant CMB anisotropy observations.  
These window functions 
are important in analyzing anisotropy measurements because
the small-scale experiments do not measure enough of the sky to produce
skymaps like COBE.  Rather they yield a few 
``band-power'' measurements of rms temperature anisotropy which reflect 
a convolution over the range of multipoles contained in the window 
function of each band.  Some instruments can produce limited 
skymaps \citep{whiteb95}.  The window function $W_\ell$ shows
how the total power observed is sensitive to the anisotropy on 
the sky as a function of angular scale:

\begin{equation}
Power = \frac{1}{4 \pi} \sum_\ell (2 \ell + 1)C_\ell W_\ell = \frac{1}{2}
(\Delta T/T_{CMB})^2 \sum_\ell \frac {2 \ell + 1}{\ell(\ell+1)} W_\ell 
\end{equation}

\noindent where the COBE normalization is $\Delta T = 27.9 \mu$K and
$T_{CMB}=2.73$K \citep{bennettetal96}.  
This allows the observations of broad-band 
power to be reported as observations of $\Delta T$, and knowing the window
function of an instrument one can turn the predicted $C_\ell$ spectrum
of a model into the corresponding prediction for $\Delta T$.  
This ``band-power'' measurement 
is based on the standard definition that for a ``flat'' power spectrum,
$\Delta T = (\ell (\ell + 1) C_\ell )^{1/2}T_{CMB}/(2\pi)$ (flat
actually means that $\ell(\ell+1)C_\ell$ is constant).

The autocorrelation function for measured temperature anisotropies
is a convolution of the true expectation values for the anisotropies 
and the window function.  Thus we have \citep{whites95}

\begin{equation}
 \left \langle 
\frac{\Delta T}{T} (\hat{n}_{1} )
\frac{\Delta T}{T} (\hat{n}_{2} ) \right \rangle = 
\frac{1}{4 \pi} \sum_{\ell=1}^{\infty} (2 \ell + 1) C_\ell 
W_\ell ( \hat{n}_{1}, \hat{n}_{2}), 
\end{equation}

\noindent
where the symmetric beam shape that is typically assumed makes
$W_{\ell}$ a 
function of separation angle only.  In general, the window function 
results from a combination of the directional response of the antenna,
the beam position as a function of time, and the weighting of each 
part of the beam trajectory in producing a temperature measurement 
\citep{whites95}.  Strictly speaking, $W_\ell$ is the diagonal part of 
a filter function $W_{\ell \ell'}$ that reflects the coupling of 
various multipoles due to the non-orthogonality of the spherical 
harmonics on a cut sky and the observing strategy of the 
instrument 
\citep{knox99}.      
It is standard to assume a Gaussian beam response of width $\sigma$, 
leading to a window function 
\begin{equation}
 W_{\ell}  = \exp [ - \ell ( \ell + 1 )\sigma^{2}]. 
\end{equation} 
The low-$\ell$
cutoff introduced by a 2-beam differencing setup comes from the window
function \citep{whitess94}  
\begin{equation}
 W_{\ell}  = 2 [ 1 - P_{\ell}(\cos \theta) ] 
\exp [ - \ell ( \ell + 1 )\sigma^{2}]. 
\end{equation}

\subsection{Sample and Cosmic Variance}

The multipoles $C_\ell$ can be related to the expected
value of the spherical harmonic coefficients by 
\begin{equation}
 \langle \sum_m{a_{\ell m}^2}\rangle = (2 \ell + 1) C_\ell 
\end{equation}
since there are $(2 \ell + 1)$  $a_{\ell m}$ for each $\ell$ and each has
an expected autocorrelation of $C_{\ell}$.  In a theory such as inflation,
the temperature fluctuations follow a Gaussian distribution about 
these expected ensemble averages.  This makes the $a_{\ell m}$ Gaussian 
random variables, resulting in a $\chi^{2}_{2 \ell + 1 }$ distribution 
for $\sum_m{a_{\ell m}^2}$.  The width of this distribution leads to a 
cosmic variance in the estimated $C_\ell$ of 
$\sigma^2_{cv} = (\ell + \frac{1}{2})^{-\frac{1}{2}}C_\ell$, 
which 
is much greater for small $\ell$ than for large $\ell$ (unless $C_\ell$ is rising in a 
manner highly inconsistent with theoretical expectations).  So, although
 cosmic variance is an unavoidable source
of error for anisotropy measurements,
it is much less of a problem for small scales 
than for COBE.  

Despite our conclusion that cosmic variance is a greater concern on 
large angular scales, Figure \ref{fig:obs_cmb} 
shows a tremendous variation in the 
level of 
anisotropy measured by small-scale experiments.  Is this evidence
for a non-Gaussian cosmological model such as topological
defects?  Does it mean we cannot trust the data?  Neither conclusion 
is justified (although both could be correct) because we do in 
fact expect a wide variation among these measurements due to their
coverage of a very small portion of the sky.  Just as it is difficult to 
measure the $C_{\ell}$ with only a few $a_{\ell m}$, 
it is challenging to 
use a small piece of the sky to measure multipoles whose spherical
harmonics cover the sphere.  It turns out that 
limited sky coverage leads to a sample variance for a particular 
multipole related to 
the cosmic variance for any value of $\ell$ by the simple formula 
\begin{equation}
 \sigma^{2}_{sv} \simeq \left ( \frac {4 \pi}{\Omega} \right ) 
\sigma^{2}_{cv}, 
\end{equation} 
where $\Omega$ is the solid angle observed \citep{scottsw94}.  One caveat: 
in testing cosmological models, this cosmic and sample variance should 
be derived from the $C_\ell$ of the model, not the observed value of the 
data.  The difference is typically small but will bias the analysis of  
forthcoming high-precision observations if cosmic and sample variance 
are not handled properly.

\subsection{Binning CMB data}

Because there are so many measurements and the most important ones have 
the smallest error bars, it is preferable to plot the data in some way that 
avoids having the least precise measurements dominate the plot.  
Quantitative analyses should weight each datapoint by the inverse of its 
variance.  Binning the data can be useful  
for display purposes but is dangerous for analysis, 
because a statistical analysis 
performed on the binned datapoints will give different results from 
one performed on the raw data.  The distribution 
of the binned errors is non-Gaussian even if the original points had 
Gaussian errors.  Binning might improve a quantitative analysis
 if the points at a particular 
angular scale showed a scatter larger than is consistent with their error 
bars, leading one to suspect that the errors have been underestimated.  
In this case, one could use the scatter to create a reasonable uncertainty on 
the binned average.  For the current CMB data there is no 
clear indication of scatter inconsistent with the errors so this is 
unnecessary.  

If one wishes 
to perform a model-dependent analysis of the data, the simplest 
reasonable approach is to 
compare the observations 
with the broad-band power estimates that should have been produced given 
a particular theory  
(the theory's $C_\ell$ are not constant so the  
window functions must be used for this).   
Combining full raw datasets is superior but computationally 
intensive (see \citealt{bondjk98a}).  A first-order correction for the 
non-gaussianity of the 
likelihood function of the band-powers has been calculated by 
\citet{bondjk98b} and is available at 
http://www.cita.utoronto.ca/\~{}knox/radical.html.

\section{Combining CMB and Large-Scale Structure Observations}

\begin{figure}
\begin{center}
\scalebox{0.5}{\includegraphics{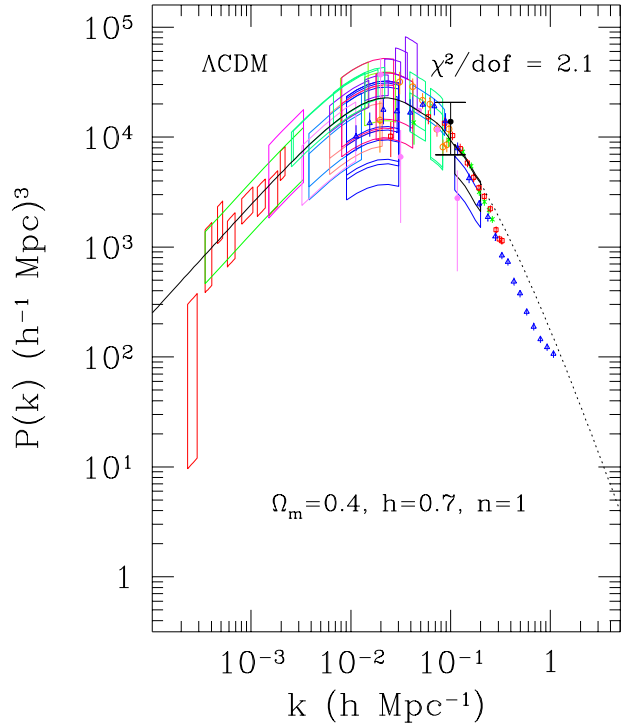}}
\caption{\small 
Compilation of CMB anisotropy detections (boxes) and large-scale structure 
observations (points with error bars) 
compared to theoretical predictions of standard $\Lambda$CDM model. 
Height of boxes (and error bars) represents
$1\sigma$ uncertainties
and width of boxes shows the full width 
at half maximum of each instrument's window function.}
\label{fig:lcdm}
\end{center}
\end{figure}

As CMB anisotropy is detected on smaller angular scales and large-scale 
structure surveys extend to larger regions, there is an increasing overlap 
in the spatial scale of inhomogeneities probed by these complementary 
techniques.  This allows us to test the gravitational instability paradigm 
in general and then move on to finding cosmological models which can 
simultaneously explain the CMB and large-scale structure observations.  
Figure \ref{fig:lcdm} shows this comparison for our compilation of CMB 
anisotropy observations (colored boxes) and of large-scale structure 
surveys (APM - \citealt{gaztanagab98}, LCRS - \citealt{linetal96}, 
Cfa2+SSRS2 - \citealt{dacostaetal94}, PSCZ - \citealt{tadrosetal99}, 
APM clusters - \citealt{tadrosed98}) including  
measurements of the dark matter fluctuations 
from peculiar velocities \citep{kolattd97} and the abundance 
of galaxy clusters \citep{vianal96,bahcallfc97}.  Plotting CMB anisotropy 
data as measurements of the matter power spectrum is a model-dependent 
procedure, and the galaxy surveys must be corrected for redshift distortions, 
non-linear evolution, and galaxy bias (see \citealt{gawisers98} 
for detailed methodology.)   Figure \ref{fig:lcdm} is good evidence 
that the matter and radiation inhomogeneities had a common origin - the 
standard $\Lambda$CDM model with a Harrison-Zel'dovich primordial power 
spectrum predicts both rather well.  On the detail level, however, the model 
is a poor fit ($\chi^2$/d.o.f.=2.1), and no cosmological model which 
is consistent with the recent Type Ia supernovae results 
fits the data much better.  Future observations will tell us if this is 
evidence of systematic problems in large-scale structure data or a fatal 
flaw of the $\Lambda$CDM model.

\section{Conclusions}

The CMB is a mature subject. The spectral distortions are well 
understood, and the Sunyaev-Zeldovich effect provides a unique tool for
studying galaxy clusters at high redshift. Global distortions will
eventually be found, most likely first at very large $l$  due to the
cumulative contributions from hot gas heated by radio galaxies, AGN, and
galaxy groups and clusters. For gas at $\sim 10^6 - 10^7$ K,
appropriate to gas in galaxy potential wells, the thermal and kinematic
contributions are likely to be comparable.

CMB anisotropies are a rapidly developing field, since the 1992
discovery with the COBE DMR of large angular scale temperature
fluctuations. At the time of writing, the first acoustic peak is being
mapped with unprecedented precision that will enable definitive
estimates to be made of the curvature parameter. More information will
come with all-sky surveys to higher resolution (MAP in 2000, PLANCK in 2007) 
that will enable most of the cosmological parameters to be
derived to better than a few percent precision if the 
adiabatic CDM paradigm proves correct. Degeneracies remain in
CMB parameter extraction, specifically between $\Omega_0$, $\Omega_b$ and
$\Omega_{\Lambda}$, but these can be removed via large-scale structure
observations, which effectively constrain $\Omega_{\Lambda}$ via weak
lensing. The goal of studying reionization will be met by the
interferometric surveys at very high resolution ($l\sim 10^3 - 10^4$).

Polarization presents the ultimate challenge, because the foregrounds
are poorly known. Experiments are underway to measure polarization at
the 10 percent level, expected on degree scales in the most optimistic
models. However one has to measure polarisation at the 1 percent level to
definitively study the ionization history and early tensor mode
generation in the universe, and this may only be possible with long
duration balloon or space experiments.

CMB anisotropies are a powerful probe of the early universe. Not only
can one hope to extract the cosmological parameters, but one should be
able to measure the primordial power spectrum of density fluctuations
laid down at the epoch of inflation, to within the uncertainties
imposed by cosmic variance. In combination with new generations of
deep wide field galaxy surveys, it should be possible to unambiguously
measure the shape of the predicted peak in the power spectrum, and
thereby establish unique constraints on the origin of the large-scale
structure of the universe.

\small

%\nocite{*}
%\bibliographystyle{plain}
\bibliographystyle{apj}
%\bibliography{uctest}
\bibliography{apj-thesis,refs}  %apj-thesis 

\end{document}